%% file: main.tex
\documentclass[runningheads]{llncs}

\usepackage[T1]{fontenc}
\usepackage{graphicx}
\usepackage{xcolor}
\usepackage{booktabs}
\usepackage{multirow}
\usepackage{amsmath}
\usepackage{amsfonts}
\usepackage{url}
\usepackage{float}
\usepackage{makecell}
\usepackage{placeins}

\begin{document}

\title{PhotonZip: Lifecycle-Aware Compression for HDF5-Based Synchrotron Workflows}
\titlerunning{PhotonZip for HDF5-Based Synchrotron Workflows}

\author{
Haoxu Li\inst{1,2} \and
Wenjing Huang\inst{1,2} \and
Chengyu Liu\inst{2,3} \and
Shiyuan Fu\inst{3} \and
Hao Hu\inst{3} \and
Guangming Tan\inst{1} \and
Dingwen Tao\inst{1}
}

\authorrunning{H. Li et al.}

\institute{
Institute of Computing Technology, Chinese Academy of Sciences
\\
\and
University of Chinese Academy of Sciences\\
\and
Institute of High Energy Physics, Chinese Academy of Sciences\\
\email{taodingwen@ict.ac.cn}
}

\maketitle

\begin{abstract}
Synchrotron detectors generate data at rates that challenge storage and I/O systems. Compression can reduce data movement, but codec-only measurements do not capture the end-to-end behavior of HDF5 workflows. We present \textbf{PhotonZip}, a lifecycle-aware compression framework that tests candidate modes through a common HDF5 path and selects among eligible measured configurations according to either compression ratio or round-trip throughput. Its modular filters preserve standard HDF5 access when the required plugin is available. Across four beamline-derived datasets, PhotonZip reaches lossless compression ratios up to 1.629$\times$ and HDF5 write throughput up to 3.40GB/s write and 7.75GB/s read. The throughput-oriented configuration achieves 1.529$\times$ lossless data reduction while improving HDF5 write and read throughput by 21.2\% and 15.0\%, respectively. Lossy preview modes provide up to 16$\times$ data reduction and reduce preview latency by up to 8.35~s in the evaluated file-per-rank HDF5 configuration.

\keywords{Scientific data compression \and HDF5 filters \and Synchrotron data \and MPI}
\end{abstract}

\input{tex/01_intro}
\input{tex/02_background}
\input{tex/03_design}
\input{tex/05_evaluation}
\input{tex/06_related_work}
\input{tex/07_conclusion}

% Add acknowledgments and disclosure statements for the camera-ready version
% as required by Springer and the conference.
\begin{credits}
\subsubsection{\ackname}
This work was supported by the National Natural Science Foundation of China through the Excellent Young Scientists Fund Program (Overseas) (Grant No. E401470), the National Key Research and Development Program of China (Grant Nos. 2020YFA0907000 and 2025YFB30037002), the National Natural Science Foundation of China (Grant Nos. 62032023 and T2125013), and Ant Group. The AI-driven experiments, simulations, and model training were performed on the robotic AI-Scientist platform of the Chinese Academy of Sciences.
% \subsubsection{\ackname} \todo{Funding and acknowledgments.}
% \subsubsection{\discintname} The authors have no competing interests to
% declare that are relevant to the content of this article.
\end{credits}

\bibliographystyle{splncs04}
\bibliography{references}

\end{document}

%% file: tex/01_intro.tex
\section{Introduction}
\label{sec:introduction}

Modern synchrotron light sources employ high-frame-rate detectors that continuously generate large volumes of scientific images~\cite{lclsii}. These data are commonly represented as multi-byte integer arrays and persisted as chunked HDF5 datasets for online processing, downstream analysis, and long-term archival~\cite{donath2023eiger2,konnecke2015nexus,marone2017tomcat}. As detector resolution, acquisition rate, and the number of beamlines continue to increase, data production is growing faster than the available storage capacity and I/O bandwidth.  Data compression is therefore important not only for reducing storage consumption, but also for decreasing the volume of data transferred through acquisition and storage pipelines. 

Applying compression in a production HDF5 workflow, however, is more difficult than selecting a codec with high standalone throughput. First, codec performance on in-memory buffers does not directly translate into end-to-end HDF5 performance, which is jointly affected by compression, data transformation, chunk and metadata management, synchronization, and storage I/O~\cite{jin2022parallelhdf5}. Second, a higher compression ratio often requires a more complex compression pipeline, introducing additional computation and reducing throughput. Compression ratio and throughput must therefore be considered as a workload-dependent trade-off rather than optimized independently. Third, comparative studies show that the relative effectiveness of compressors and preprocessing pipelines can change across datasets and operating configurations~\cite{bernstein2024lossless,zhao2020sdrbench}. Finally, deploying compression in an actual scientific workflow requires users to understand codec-specific parameters, manually compare alternatives, integrate them with HDF5, and preserve sufficient metadata for subsequent decompression. These requirements impose a substantial configuration and integration burden on beamline users.

To address these challenges, we present \textbf{PhotonZip}, a lifecycle-aware compression system for HDF5-based synchrotron data workflows. Rather than relying on one codec for every data product, PhotonZip maps workflow intent to complementary compression paths: lossy compression for online processing when information loss is acceptable and lossless compression for persistent archival. PhotonZip tests candidate modes through a common HDF5 procedure and uses hard eligibility constraints followed by a single-objective decision rule to select a measured configuration. Codec-specific filter properties and payload headers preserve the parameters required for decoding, allowing applications to retain standard HDF5 read and write calls when the corresponding plugin is installed. We further characterize the complete I/O path using file-per-rank MPI execution and backend threading rather than treating compression as an isolated kernel.

% We evaluate PhotonZip using \needdata{number and names of representative beamline-derived datasets} on \needdata{target computing platforms}, comparing it with raw HDF5 and \needdata{representative HDF5 compression baselines}. The evaluation examines storage reduction, end-to-end HDF5 read/write throughput, configuration selection, and process-level scalability under a consistent measurement methodology. PhotonZip achieves \needdata{...}

In summary, this paper makes the following contributions:

\begin{itemize}
    \item \textbf{Transparent compression for HDF5 workflows.}
    We integrate complementary lossless and lossy backends as HDF5 filters whose identifiers, parameters, and payload headers support standard HDF5 reads and writes when the required plugins are available.

    \item \textbf{Goal-directed compression configuration.}
    We develop a lightweight configuration mechanism that tests candidate modes through a common HDF5 procedure, filters the measured candidates by fidelity and data compatibility, and selects the best configuration for the user's ratio or throughput objective.

    \item \textbf{Modular and extensible implementation.}
    We provide modular interfaces that separate compression backends, preprocessing strategies, configuration policies, and HDF5 integration, allowing additional compression algorithms to be incorporated without redesigning the end-to-end data path.

    \item \textbf{End-to-end workflow characterization.}
    We evaluate PhotonZip using beamline-derived datasets and a file-per-rank MPI benchmark, quantifying compression effectiveness and complete HDF5 I/O performance across process--thread configurations.
\end{itemize}

%% file: tex/02_background.tex
\section{Background and Gap Analysis}
\label{sec:background}

\subsection{Synchrotron Data Workflows}
\label{sec:background-workflow}

Many synchrotron imaging experiments use high-frame-rate detectors to acquire sequences of multidimensional images. The resulting multi-byte integer arrays move through data acquisition to online processing, persistent storage, and downstream analysis~\cite{donath2023eiger2,marone2017tomcat}. Their distributions, noise levels, and temporal correlations vary across detectors and experiments, producing dataset-dependent compression behavior~\cite{bernstein2024lossless}.
HDF5 stores multidimensional arrays and metadata and underlies scientific formats such as NeXus~\cite{folk2011hdf5,konnecke2015nexus}. Its chunked filter pipeline can compress each chunk during a write and decompress it on demand during a read~\cite{hdfgroupfilters}. Applications retain standard HDF5 interfaces when the required filter is available, while the file records the filter and parameters needed for decoding.

\begin{figure}[t]
    \centering
    \includegraphics[width=0.72\linewidth]{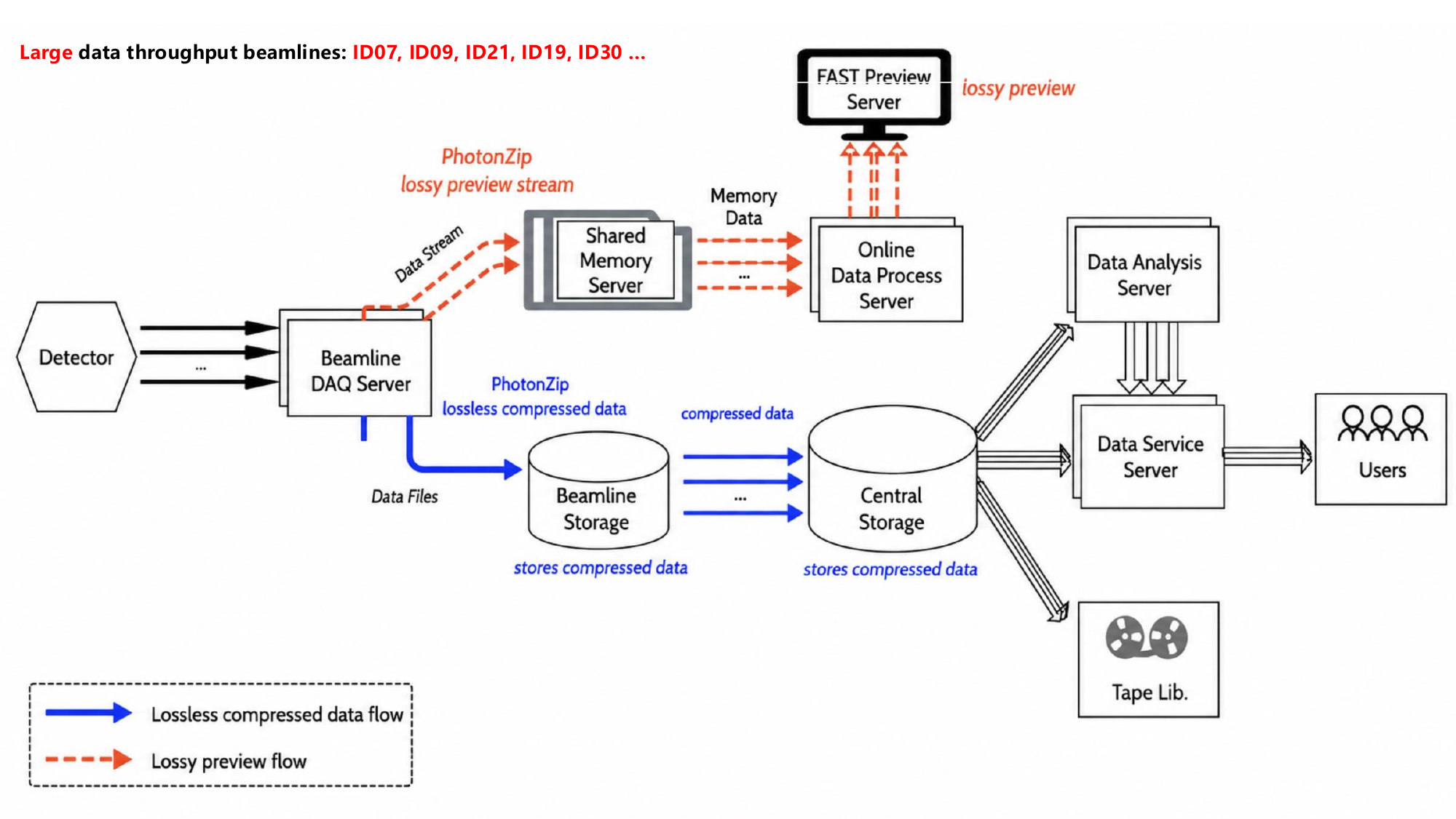}
    \caption{PhotonZip's integration point in a synchrotron data workflow.}
    \label{fig:workflow-integration}
\end{figure}

Figure~\ref{fig:workflow-integration} shows the target integration point. PhotonZip is placed between data acquisition and HDF5 storage to reduce downstream data movement without changing surrounding HDF5 interfaces. When information loss is acceptable, it can generate a lossy product for online processing; the archival branch instead applies lossless compression directly to the acquired data.

\subsection{Gap Analysis and Design Goals}
\label{sec:background-gap}

\paragraph{Standalone codec performance does not represent HDF5 performance.}
In-memory compression ratio and throughput do not determine whether compression benefits an HDF5 workflow. Data conversion, filter execution, synchronization, metadata, and storage behavior also contribute to end-to-end performance~\cite{jin2022parallelhdf5}; parallel execution additionally depends on work partitioning and process configuration.

\paragraph{A fixed configuration cannot satisfy different workflow objectives.}
Compression methods occupy different points in the ratio--throughput--fidelity space. PhotonZip currently integrates MANS and LC for lossless storage and PFPL and spatial downsampling for lossy preview; their different data types and operating characteristics motivate lifecycle-aware selection. Online products may use either lossy mode when information loss is acceptable, while archival products require lossless compression and may prioritize either ratio or throughput~\cite{huang2025mans,azami2025lc}.

\paragraph{Codec-specific integration creates deployment burden.}
HDF5 filters and scientific compressors expose heterogeneous choices such as filter availability, pipeline composition, compression level, and lossy quality setting~\cite{hdfgroupfilters,azami2025lc,li2023sperr}. Users must translate workflow requirements into these settings and ensure that the corresponding filter remains available for decoding. Tightly coupling this logic to HDF5 access also makes codecs and policies harder to replace or extend.

Based on these gaps, PhotonZip has four design goals:
\begin{itemize}
    \item \textbf{G1: Transparent HDF5 integration.} Preserve standard HDF5 reads and writes through installed filters and retain the codec-specific parameters required for decoding.
    \item \textbf{G2: Lifecycle-aware configuration.} Test candidate modes through a common HDF5 procedure, filter them by workflow and fidelity requirements, and select for compression ratio or throughput.
    \item \textbf{G3: Modular extensibility.} Separate backends, preprocessing, configuration, and HDF5 integration so compatible algorithms can be added without redesigning the data path.
    \item \textbf{G4: End-to-end characterization.} Evaluate file-per-rank HDF5 execution with backend threading across process--thread configurations.
\end{itemize}

%% file: tex/03_design.tex
\section{PhotonZip Design}
\label{sec:design}

\subsection{System Overview}
\label{sec:design-overview}

PhotonZip bridges workflow-level requirements and codec-specific execution through four components: a lightweight configuration procedure, modular compression backends, codec-specific HDF5 filters, and an end-to-end evaluation layer. Users specify a dataset's purpose, fidelity requirement, and optimization objective. PhotonZip tests eligible modes through a common HDF5 procedure and applies a deterministic rule to choose a measured configuration. Compression and decompression then operate through the selected HDF5 filter without requiring application code to invoke codec kernels directly.

The key observation is that synchrotron data products can have different requirements at different lifecycle stages. Archival products require lossless compression, whereas an online consumer may explicitly permit lossy compression to reduce data movement. PhotonZip treats this fidelity choice as a hard constraint before comparing measured candidates by the requested objective. If both products are required, they must be generated as separate branches from the acquired data; applying a lossless codec to lossy reconstructed data does not create a lossless archive.

\subsection{Transparent HDF5 Data Path}
\label{sec:design-hdf5}

PhotonZip treats HDF5 as a data contract: applications access logical data arrays through HDF5 calls, while installed filters manage their physical compressed representation. The HDF5 chunk is the encoding boundary. On a write, the selected filter transforms a logical chunk into an encoded buffer; on a read, it applies the inverse path before returning the chunk to the application~\cite{folk2011hdf5,hdfgroupfilters}.

HDF5 records the filter identifier and its client parameters in the dataset creation properties, while each backend payload stores additional fields needed by its decoder, such as the original chunk size. These mechanisms preserve codec-specific decoding information, but they do not remove the deployment dependency: a compatible filter plugin must be installed on the reader. The standalone PhotonZip container additionally records array shape, data type, backend parameters, and preprocessing information; this richer container metadata is separate from the current HDF5 filter paths.

\subsection{Lifecycle-Aware Goal-Directed Configuration}
\label{sec:design-selection}

PhotonZip accepts a compact request consisting of data purpose, a lossless or lossy fidelity choice, the ratio or throughput objective, and an optional quality preset for lossy compression. It first filters the available modes by hard constraints, including fidelity, data type, dimensionality, and backend availability. Lossless requests admit only exactly reconstructable modes, while lossy modes are eligible only when the user permits information loss. Fidelity is therefore established before performance is compared.

All eligible modes are tested on the same representative HDF5 sample under a common measurement procedure. Let $S$ denote its logical data, $B_{\mathrm{HDF5}}(c,S)$ the resulting HDF5 storage size, and $\mathcal{C}_{\mathrm{valid}}$ the modes that pass the hard constraints. PhotonZip measures
\begin{equation}
CR(c,S)=\frac{|S|}{B_{\mathrm{HDF5}}(c,S)},\qquad
Q_{\mathrm{RT}}(c,S)=\frac{2|S|}{T_{\mathrm{write}}(c,S)+T_{\mathrm{read}}(c,S)},
\label{eq:hdf5-selection-throughput}
\end{equation}
where write and read times cover the complete HDF5 path, including filtering, chunk handling, synchronization, and storage I/O. The user requirement then determines the measured winner:
\begin{equation}
c^*=\begin{cases}
\arg\max_{c\in\mathcal{C}_{\mathrm{valid}}} CR(c,S),
& \text{ratio priority},\\
\arg\max_{c\in\mathcal{C}_{\mathrm{valid}}} Q_{\mathrm{RT}}(c,S),
& \text{throughput priority}.
\end{cases}
\label{eq:selection-policy}
\end{equation}
This single-objective decision requires neither a weighted score nor a Pareto search. The selected mode and parameters are fixed for the complete dataset rather than changed independently for individual chunks.

\subsection{Modular Compression Architecture}
\label{sec:design-modular}

\begin{figure}[t]
    \centering
    \includegraphics[width=0.65\linewidth]{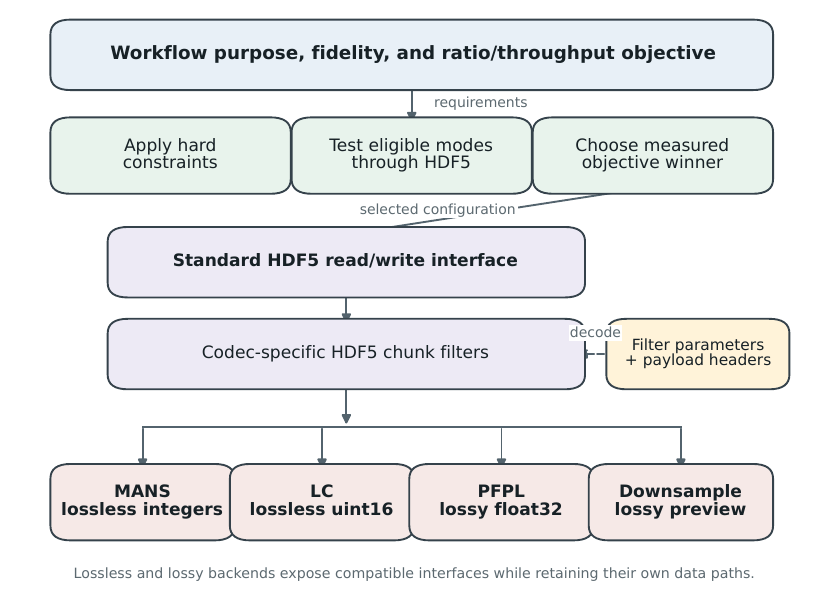}
    \caption{PhotonZip's modular organization.}
    \label{fig:modular-architecture}
\end{figure}

Figure~\ref{fig:modular-architecture} shows PhotonZip's HDF5-oriented modular path. After configuration selection, applications retain standard HDF5 read and write calls, while codec-specific chunk filters dispatch data to the selected backend. Filter parameters and payload headers preserve the state required for decoding. Backend data types, parameters, threading, and encoded formats remain encapsulated, so codecs compatible with the filter interface can be added without redesigning the upper layers.

The current implementation integrates MANS for lossless multi-byte integers, LC for lossless 16-bit unsigned data, and two lossy preview modes: PFPL and spatial downsampling. PFPL operates on single-precision data, while downsampling reduces the spatial data volume before reconstruction. Both lossy modes are candidates for the G2 selection procedure and can be compared by compression ratio or end-to-end HDF5 throughput. A separate CLI prototype optionally applies whole-volume inter-frame delta to 16-bit unsigned arrays along axis~0. It computes differences in a widened type and clips them to signed 16-bit values; reconstruction is exact only when no clipping occurs. This prototype is not part of the current HDF5 filters.

\subsection{Hybrid Process--Thread Execution}
\label{sec:design-parallel}

PhotonZip adopts a file-per-rank execution model. Each MPI rank processes a rank-local data slab, applies the same selected mode, and reads or writes a private HDF5 file. HDF5 invokes the selected filter independently for each chunk, so codec execution requires no communication between ranks. MPI coordinates the ranks and aggregates their results.

Process-level concurrency is combined with backend-internal OpenMP parallelism. For a fixed CPU budget, PhotonZip assigns approximately $P/R$ threads to each of $R$ ranks, where $P$ is the number of available cores. Overall throughput is computed as
\begin{equation}
Q=\frac{\sum_{r=1}^{R}|D_r|}{\max_{r} T_r},
\label{eq:parallel-throughput}
\end{equation}
where $|D_r|$ and $T_r$ are the logical source bytes and end-to-end HDF5 time of rank $r$. This model balances independent HDF5 streams against compression threads per stream while keeping the selected mode unchanged across ranks.

%% file: tex/05_evaluation.tex
\section{Evaluation}
\label{sec:evaluation}
% 本节围绕以下问题组织实验：
% RQ1：PhotonZip 相对现有 HDF5 Filters 的压缩率--吞吐折中如何？
% RQ2：目标驱动配置能否针对不同目标选择接近最佳的候选？
% RQ3：PhotonZip 的 MPI 扩展性和进程/线程配置影响如何？
% RQ4：delta 和 HDF5 chunk size 对性能有什么影响？
% RQ5：PhotonZip 在同步辐射源数据通路中能取得什么端到端收益？

\subsection{Experimental Setup}
\label{sec:evaluation-methodology}

We conduct all experiments on a dual-socket Intel Xeon Gold 6530
server. The detailed hardware configuration is summarized in
Table~\ref{tab:hardware_setup}.

\begin{table}[H]
\centering
\caption{Hardware configuration of the experimental platform.}
\label{tab:hardware_setup}
\renewcommand{\arraystretch}{1.08}
\begin{tabular}{@{}p{0.27\linewidth}p{0.55\linewidth}@{}}
\toprule
\textbf{Component} & \textbf{Specification} \\
\midrule
Processor    & 2 $\times$ Intel Xeon Gold 6530 \\
CPU cores    & 32 cores per socket; 64 physical cores in total \\
CPU threads  & 1 thread per core; 64 logical CPUs \\
NUMA         & 2 NUMA nodes \\
Vector ISA   & AVX-512 \\
Memory       & 503 GB \\
Storage      & 3.5 TB NVMe SSD \\
\bottomrule
\end{tabular}
\end{table}

\subsection{Compression and HDF5 Performance}
\label{sec:evaluation-compression}

\begin{figure}[!ht]
    \centering
    \includegraphics[width=0.7\linewidth]{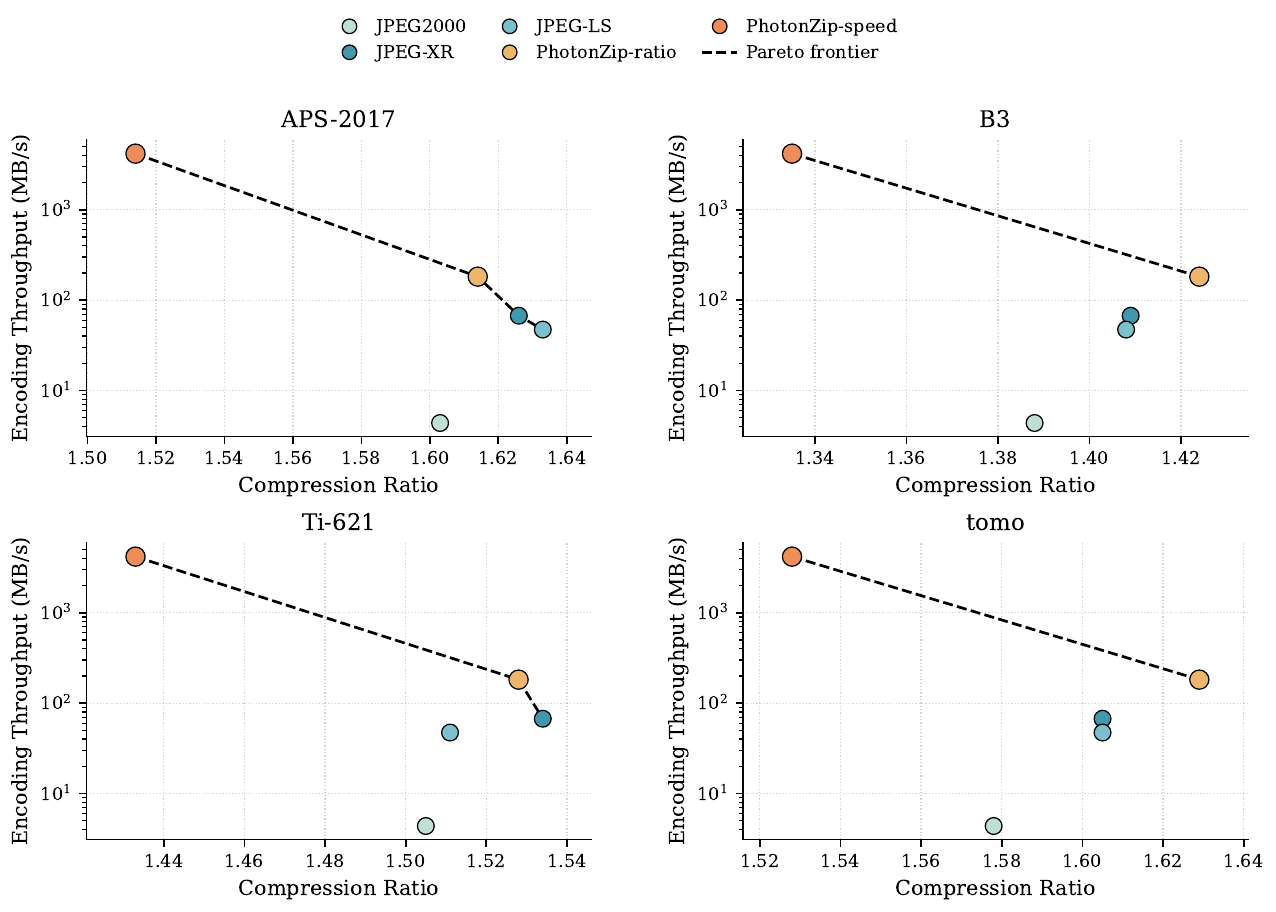}
    \caption{Compression-ratio--throughput trade-off of lossless
    methods across four datasets. Dashed lines indicate the Pareto
    frontier.}
    \label{fig:ratio-speed-pareto}
\end{figure}

Figure~\ref{fig:ratio-speed-pareto} compares the lossless
compression operating points across four datasets.

PhotonZip-Ratio consistently favors stronger compression, reaching
1.424$\times$--1.629$\times$ across the four datasets, whereas
PhotonZip-Speed trades a small amount of compression ratio for
substantially higher codec throughput. The two configurations thus
cover complementary lossless operating points for storage- and
throughput-oriented workloads.

We include a $32\times2$ configuration to examine a decomposition aligned
with the two NUMA domains. The process--thread results below show that
the best decomposition depends on the selected configuration.

We further evaluate these operating points in the end-to-end HDF5
path. Table~\ref{tab:hdf5-summary} also includes the two integrated
lossy preview modes, PhotonZip-PFPL and PhotonZip-Downsample, which are
additional candidates for goal-directed selection and provide substantially
higher data reduction.

To examine the effect of process--thread decomposition, we further
compare representative PhotonZip configurations under different
parallel settings. Table~\ref{tab:photonzip-parallel} reports the
resulting end-to-end HDF5 throughput.

\begin{table}[htbp]
\centering
\caption{PhotonZip HDF5 throughput under different configurations (MiB/s).}
\label{tab:photonzip-parallel}

\begin{tabular}{@{}llll@{}}
\toprule
\textbf{Method} &
\textbf{Ranks $\times$ Threads} &
\textbf{Write} &
\textbf{Read} \\
\midrule
PhotonZip-Ratio & $1 \times 64$ & 202  & 176  \\
                 & $8 \times 8$  & 993  & 1025 \\
                 & $16 \times 4$ & 1374 & 1482 \\
                 & $32 \times 2$ & 1995 & 2062 \\
\midrule
PhotonZip-Speed & $32 \times 2$ & 3399 & 7573 \\
                 & $64 \times 1$ & 3110 & 4020 \\
\bottomrule
\end{tabular}
\end{table}

PhotonZip-Ratio benefits strongly from increased process-level
parallelism up to 32 ranks. For PhotonZip-Speed, the $32\times2$
configuration outperforms $64\times1$, indicating that the optimal
process--thread decomposition depends on the system configuration and
backend execution behavior.

\begin{table}[htbp]
\centering
\caption{End-to-end HDF5 performance across PhotonZip operating points.}
\label{tab:hdf5-summary}
\begin{tabular}{@{}lllll@{}}
\toprule
\textbf{Method} &
\textbf{Data Reduction} &
\textbf{Write (MiB/s)} &
\textbf{Read (MiB/s)} &
\textbf{PSNR} \\
\midrule
Raw HDF5
& 1.00$\times$
& 2804
& 6586
& -- \\

PhotonZip-Ratio
& 1.629$\times$
& 2615
& 3664
& Lossless \\

PhotonZip-Speed
& 1.529$\times$
& 3399
& 7573
& Lossless \\

PhotonZip-PFPL
& 15.96$\times$
& 3956
& 6814
& 40.2 dB \\

PhotonZip-Downsample
& 16.00$\times$
& 60435
& 29664
& 45.45 dB \\
\bottomrule
\end{tabular}
\end{table}

PhotonZip-Speed improves both HDF5 write and read throughput over raw
I/O while retaining lossless compression. The lossy operating points
provide much larger data reduction, with PFPL reaching 15.96$\times$
and PhotonZip-Downsample achieving a similar reduction at higher
reconstruction quality. Together, these results show that
PhotonZip spans a broad trade-off space from lossless high-throughput
I/O to aggressive lossy data reduction.

\subsection{Ablation of Goal-Directed Configuration}
\label{sec:evaluation-ablation}

We compare PhotonZip with fixed MANS and LC configurations across
four datasets. PhotonZip-Ratio targets compression effectiveness,
while PhotonZip-Speed targets throughput.

\begin{figure}[!htbp]
    \centering
    \includegraphics[width=0.8\linewidth]{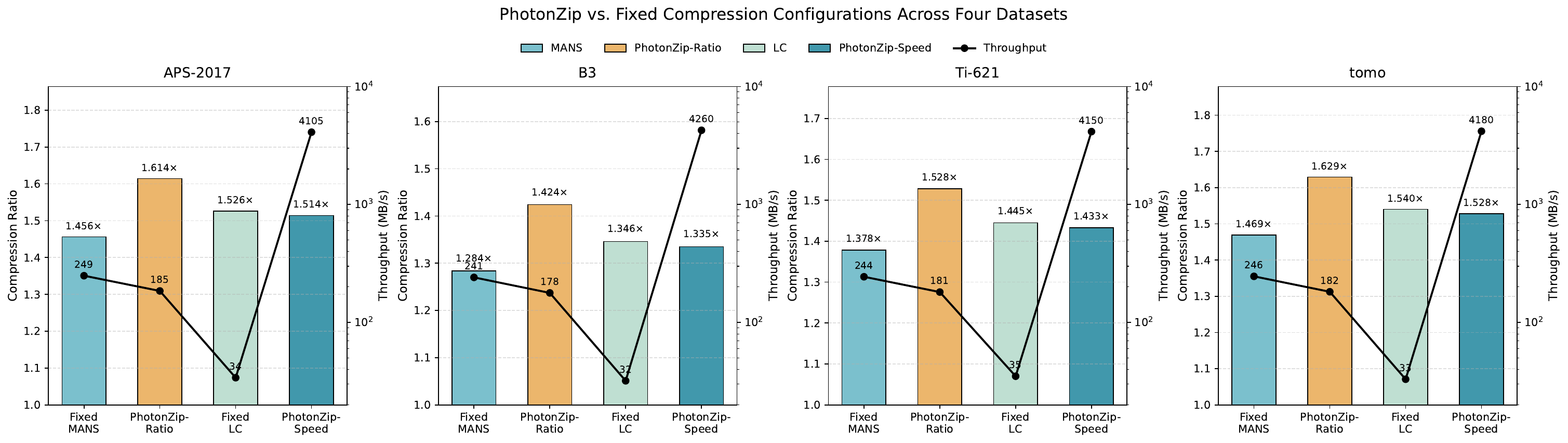}
    \caption{Comparison of PhotonZip with fixed compression
    configurations.}
    \label{fig:photonzip-vs-fixed}
\end{figure}
\FloatBarrier

PhotonZip-Ratio consistently improves compression ratio over fixed
MANS. On \textit{tomo}, the ratio increases from 1.469$\times$ to
1.629$\times$. In contrast, PhotonZip-Speed selects the throughput-oriented
configuration and sustains up to 4.18~GB/s end-to-end HDF5 throughput
with only a small reduction in compression ratio. These results show
that goal-directed configuration enables PhotonZip to adapt to
different compression and throughput objectives.

\subsection{Synchrotron Workflow Integration}
\label{sec:evaluation-workflow}
% 回答 RQ5，介绍 PhotonZip 接入 detector/DAQ--HDF5--storage 数据流的方式。
% 如果是真实 beamline 部署，报告持续运行时间、输入速率、backlog、丢帧、资源占用和存储节省。
% 如果是 TIFF 回放或本地模拟，应明确称为 production-inspired 或 emulated workflow，不称为生产部署。
% 对比 raw、代表性通用压缩和 PhotonZip 的端到端处理能力。

We finally evaluate PhotonZip for fast preview in the synchrotron
workflow. Figure~\ref{fig:preview-quality} compares the reconstructed
data produced by PhotonZip-PFPL and PhotonZip-Downsample with the original data.

The end-to-end preview time is reduced from 42.33~s for raw data to 39.44~s with PhotonZip-PFPL (2.89~s reduced) and 33.98~s with
PhotonZip-Downsample (8.35~s reduced). Since visualization alone
takes approximately 33~s, the remaining data-processing overhead is
reduced from 9.33~s to 6.44~s with PhotonZip-PFPL and to only 0.98~s with
PhotonZip-Downsample. These results show that PhotonZip accelerates scientific
preview by reducing the data movement and preparation cost while
preserving sufficient reconstruction quality for rapid inspection.

\FloatBarrier

%% file: tex/06_related_work.tex
\section{Related Work}
\label{sec:related-work}

\subsubsection{Scientific Data Compression.}
Scientific compressors span lossless methods, which preserve exact values, and lossy methods, which trade reconstruction fidelity for greater data reduction. Lossless approaches combine entropy coding and reversible transformations to exploit structure in typed arrays; MANS targets multi-byte integers, while LC composes lightweight transformations and encoders into configurable pipelines~\cite{masui2015bitshuffle,huang2025mans,azami2025lc}. Representative lossy compressors include SZ, ZFP, MGARD, and SPERR, which use prediction, block transforms, multilevel representations, or wavelet coding~\cite{di2016sz,lindstrom2014zfp,ainsworth2019mgard,li2023sperr}. Comparative studies show that compression ratio and throughput vary across datasets and configurations~\cite{bernstein2024lossless,zhao2020sdrbench}; workflows must therefore choose among complementary methods rather than assume one fixed codec is universally preferable.

\subsubsection{Compression in HDF5 and Scientific Workflows.}
HDF5 applies compression through chunk-level filters while preserving its dataset interface~\cite{folk2011hdf5,hdfgroupfilters}. Third-party filters extend the available codecs, but applications must still select a filter, configure codec-specific parameters, and retain the implementation required for decoding. Scientific I/O systems further combine data transformation with parallel storage and movement. Studies of parallel HDF5 compression and ADIOS2 show that workflow performance includes filter execution, synchronization, metadata, layout, and physical I/O, and therefore cannot be inferred from codec-only throughput~\cite{jin2022parallelhdf5,godoy2020adios2}. These systems provide execution mechanisms for compressed I/O, but do not by themselves provide a workflow-level policy for coordinating lossless and lossy codecs.

\begin{figure}[t]
    \centering
    \includegraphics[width=0.68\linewidth]{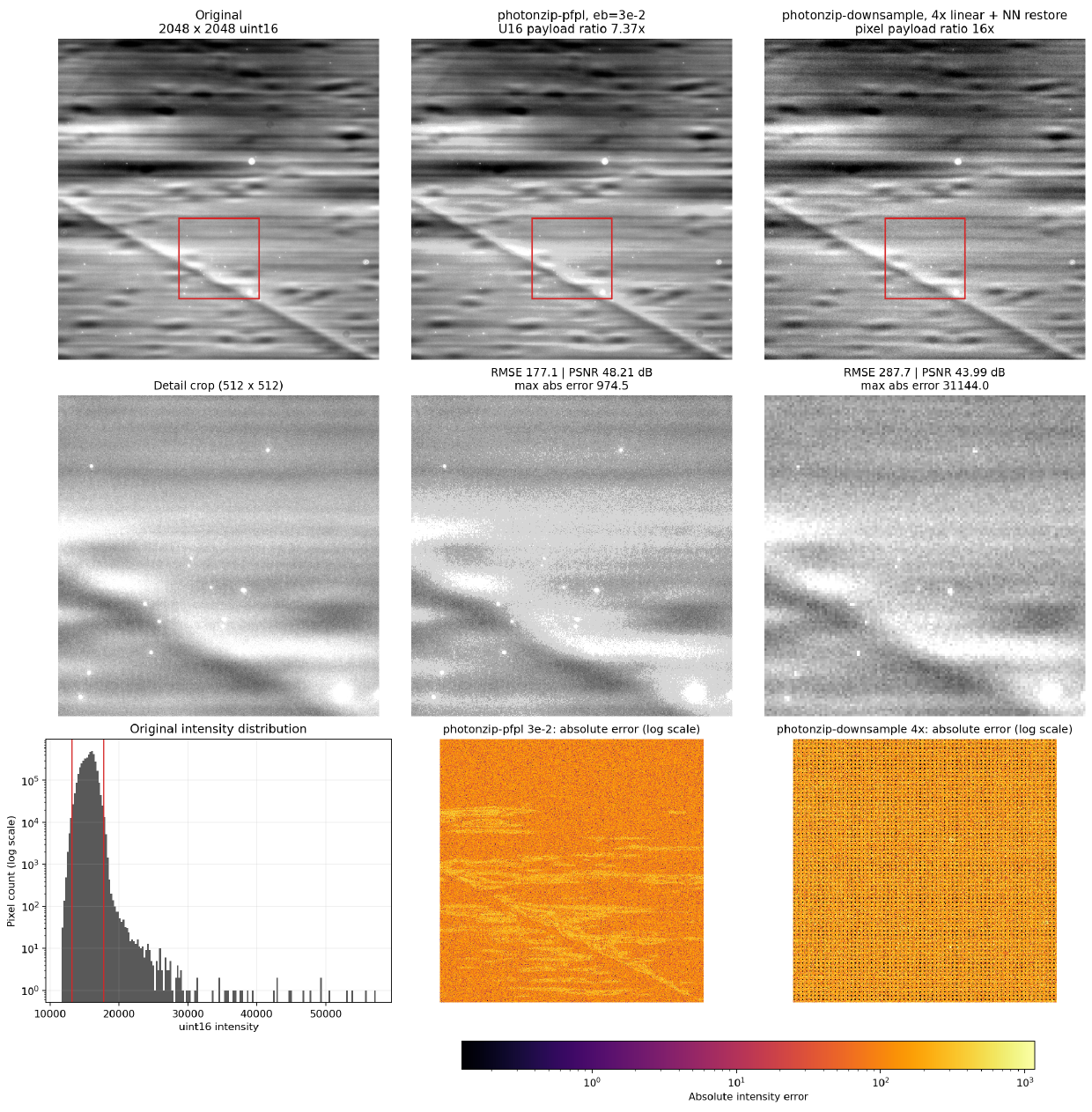}
    \caption{Reconstruction quality of PhotonZip fast-preview modes.
    PFPL and PhotonZip-Downsample provide different trade-offs between
    data reduction and visual fidelity.}
    \label{fig:preview-quality}
\end{figure}

%% file: tex/07_conclusion.tex
\section{Conclusion}
\label{sec:conclusion}

PhotonZip provides lifecycle-aware compression for HDF5-based synchrotron workflows by combining codec-specific filters, modular backends, and a fidelity-first single-objective configuration procedure. Across four beamline-derived datasets, its lossless modes reach compression ratios up to 1.629$\times$, while the throughput-oriented mode achieves 1.529$\times$ data reduction with improved end-to-end HDF5 performance. The lossy preview path provides nearly 16$\times$ data reduction and reduces preview latency by up to 8.35~s. These results are obtained on a single-node, file-per-rank HDF5 configuration and do not represent shared-file Parallel HDF5 or continuous beamline deployment.